# Computed extended depth of field optical-resolution photoacoustic microscope

Xianlin Song [a, #, *], Jianshuang Wei [b, c, #], Lingfang Song [d]

[a] School of Information Engineering, Nanchang University, Nanchang, China;
[b] Britton Chance Center for Biomedical Photonics, Wuhan National Laboratory for Optoelectronics-Huazhong University of Science and Technology, Wuhan, China;
[c] Moe Key Laboratory of Biomedical Photonics of Ministry of Education, Department of Biomedical Engineering, Huazhong University of Science and Technology, Wuhan 430074, China;
[d] Nanchang Normal University, Nanchang 330031, China;
[#] equally contributed to this work
*songxianlin@ncu.edu.cn

**Abstract**：Photoacoustic microscopy with large depth of focus is significant to the biomedical research. The conventional optical-resolution photoacoustic microscope (OR-PAM) suffers from limited depth of field (DoF) since the employed focused Gaussian beam only has a narrow depth range in focus, little details in depth direction can be revealed. Here, we developed a computed extended depth of field method for photoacoustic microscope by using wavelet transform image fusion rules. Wavelet transform is performed on the max amplitude projection (MAP) images acquired at different axial positions by OR-PAM to separate the low and high frequencies, respectively. The fused low frequency coefficients is taking the average of the low-frequency coefficients of the low-frequency part of the images. And maximum selection rule is used in high frequency coefficients. Wavelet coefficient of the MAP images are compared and select the maximum value coefficient is taken as fused high- frequency coefficients. And finally the wavelet inverse transform is performed to achieve large DoF. Simulation was performed to demonstrate that this method can extend the depth of field of PAM two times without the sacrifice of lateral resolution. And the in vivo imaging of the mouse cerebral vasculature with intact skull further demonstrates the feasibility of our method.

## 1 Introduction

Photoacoustic imaging is a new medical imaging method based on the photoacoustic effect [1]. The photoacoustic effect was proposed in 1880 by Bell. A laser pulse irradiates the sample, the sample absorbs pulse energy, then the temperature rises instantaneously, and adiabatic expansion occurs, which then generates ultrasonic signal. Because the ultrasonic signal is excited by laser pulses, it is called photoacoustic signal. The photoacoustic effect can be successfully applied to biological tissue imaging because of the large differences in the absorption of laser pulses in different biological tissues [2-4]. The photoacoustic signal received by the ultrasonic probe contains the absorption characteristics of biological tissues. The optical absorption distribution can be reconstructed by the reconstruction algorithm.

As a promising tool for biomedical research, optical-resolution photoacoustic microscopy (OR- PAM) is a noninvasive biomedical imaging technique with high-resolution [5]. Various applications has been implemented, which include structural, functional and molecular imaging of tissues [6-8]. The conventional OR-PAM employs focused gaussian beam to achieve high lateral resolution by a microscope objective with high numerical apertures, which can achieve micron- to submicron-sized focal spot. Since the focused gaussian beam only has narrow depth range in focus, little detail in depth direction can be revealed [9]. To address this issue, many methods are developed to extend the depth of field (DoF). The non-diffracting properties of bessel beam make it own large depth of field compare with gaussian beam [10], while the artifacts introduced by the side lobes of the Bessel beam should be suppressd by non-linear method. Electrically tunable lens (ETL) has also been introduced in OR-PAM [11]. This resulted in a focus-shifting time of about 15 ms. It is fast enough for pulsed lasers with a repetition rate of tens of hertz, while being quite slow for those lasers with repetition rate of kilos to hundreds of kilohertz, which has been widely used in the OR-PAM. Tunable acoustic gradient lens has also been introduced into OR-PAM [12, 13], but it needs special synchronization circuits to control the laser, which made system more complicated. Depth scanning by using motorized stage is commonly used as it is the most convenient approach [14, 15]. However, this method limits the volumetric imaging speed with its slow mechanical adjustment, and it is a challenge that the images obtained by scanning at different depth fuse into a clear image.

In this manuscript, we report a novel method to address the problem, we employ wavelet transform image fusion rules to extend the DoF, wavelet coefficient from several source images are fused by electing average of the approximation coefcients and maximum of the detailed coefcients. Fused images are obtained by taking inverse wavelet transform. Simulation experiment was carried out to demonstrate the computed extended depth of field of the method. The in vivo imaging of the mouse cerebral vasculature with intact skull further demonstrates the feasibility of our method.

## 2 METHOD

### 2.1 Fusion Rules

Image fusion process is used to associate the two or more images into one image. Fusion image are more explanatory than single source image, and more details can be revealed due to the extended DoF. Overall, it consists of three steps, as shown in Figure 1. The first step is wavelet decomposition, the max amplitude projection (MAP) images (source images) were acquired at different depth by homemade OR-PAM [13], wavelet transform was carried out on each source image. And we obtained a sequence of decomposition wavelet coefficients. The second step is fusion rules, wavelet transform separates the low and high frequencies, respectively. The fused low fequency coeffcients are obtained by several images wavelet coefficient are averaged, as shown in (1).

$$w(x, y) = 0.5 \times [w_1(x, y) + w_2(x, y) + ... + w_n(x, y)] \quad (1)$$

Where, $w_1(x, y)$, $w_2(x, y)$ is the wavelet coefficient of source image 1 and 2, respectively. n is the number of MAP images that obtained at different depth.

Wavelet coefficient with a large absolute value corresponds to a significant feature of the image. And maximum selection rule is used in high frequency coefficients. Wavelet coefficient of the MAP images are compared and select the maximum value coefficient is taken as fused high- frequency coefficients, as shown in (2)

$$w(x, y) = \max[w_1(x, y), w_2(x, y), ..., w_n(x, y)] \quad (2)$$

Finally, a new set of fused wavelet coefficients is obtained, and a high-resolution imaged with computed extended depth of field is achieved by inverse wavelet transform.

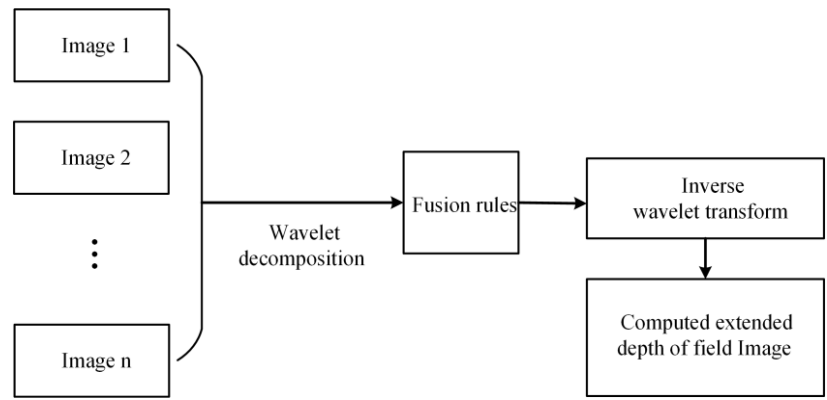

Figure 1. Wavelet based fusion rules process. The fused low fequency coeffcients are obtained by several images wavelet coefficient are averaged. And maximum selection rule is used in high frequency coefficients. Wavelet coefficient of the MAP images are compared and select the maximum value coefficient is taken as fused high- frequency coefficients. Finally, we achieved computed extended depth of field by using inverse wavelet transform.

## 3 RESULTS

### 3.1 Performance measure

Different focused images were used to verify the feasibility of the method. As shown in Figure 2, Figure 2 (a) and Figure 2 (b) are the same images with different focus. Figure 2 (a) front bottle is focused, back screen is out of focus. Figure 2 (b) front bottle is out of focus, back screen is focused. Figure 2 (c) is obtained after using the fusion rules, both the front bottles and the back screen are focused, more details can be resolved along the depth. The fused image contains more informative compare to the individual source images. Therefore, computed extended depth of field was achieved by using our method. Performance of our method was analyzed by calculating Entropy, Average gradient, Mean Square Error (MSE) and Edge strength, as shown in Table 1.

Entropy is a major indicator of the richness of information in an image, which indicates the average amount of information contained in an image. The fused image has bigger entropy value (6.6109) compare to the source image 1 (6.2642) and source image 2 (5.7876), respectively. The average gradient represents the contrast of details of image. The average gradient is also called the sharpness of the image, reflecting the contrast and texture transformation characteristics of detail. The average gradient of fused image is 3.6965, while, the average gradient of image1 and image 2 is 2.8874 and 2.1028, respectively. Standard deviation is a statistical concept that expresses the degree of dispersion, and it reflects the distribution of gray values of an image. The magnitude of the standard deviation is positively related to the quality of the image. The larger the standard deviation, the more scattered the grayscale distribution of the image, the more useful information the image has. The standard deviation of fused image is 40.6950, higher than that of image1(36.4650) and image 2 (29.0121), respectively. The edge strength was also used to evaluate the performance of our method, the edge strength of fused image is 40.9111, while, the edge strength of image1 and image 2 is 31.8601 and 23.0467, respectively. The quality of the fused image is significantly better than that of a single source image, and it contains more details, which indicates a larger (~ two times) depth of field the fused image has.

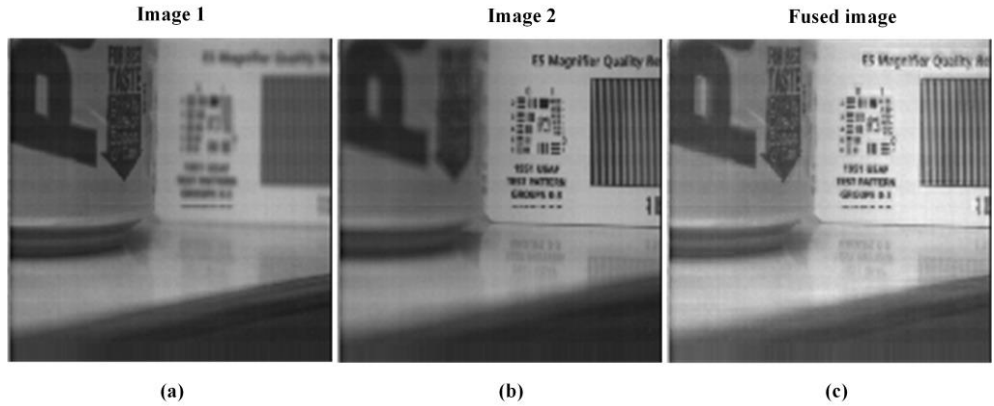

Figure 2. Simulation. (a) Source image 1, the front bottle is focused, and the back screen is out of focus. (b) Source image 2, the front bottle is out of focus, and the back screen is focused. (c) Fused Image, both the front bottles and the back screen are focused, more details can be resolved along the depth.

Table 1. Performance measure of image fusion.

|  | Entropy | Average gradient | Standard deviation | Edge strength |
|---|---|---|---|---|
| **Image 1** | 6.2642 | 2.8874 | 36.4650 | 31.8601 |
| **Image 2** | 5.7876 | 2.1028 | 29.0121 | 23.0467 |
| **Fused image** | 6.6109 | 3.6965 | 40.6950 | 40.9111 |

### 3.2 *In vivo* imaging of the mouse cerebral vasculature with intact skull

To fully demonstrate the performance of our method, we performed *in vivo* imaging of the mouse cerebral vasculature with intact skull. A 5-week-old female Balb/c mouse (Animal Biosafety Level 3 Laboratory, Wuhan, China) was used for imaging. Before imaging, the mouse was anesthetized by intraperitoneal injection of chloral hydrate (0.2 g/kg) and urethane (1 g/kg). And then the head of the mouse was fixed on a brain stereotactic apparatus for further operation. Most of the scalp and fascia above the skull was removed, forming an imaging window of $4 \times 4$ mm$^2$. Ultrasonic gel was applied for acoustic coupling. All procedures were carried out in accordance with the Institutional Animal Care and Use Committee of Hubei Province.

A homemade single focus OR-PAM system with high resolution reached to micrometer as previously described [13], the DoF of the system is ~ 120 μm. First, we set the focal plane of the system to the surface of the skull, a two-dimensional raster scanning with a step size of 3 μm is performed to acquire the depth-dependent PA signal (source PA image 1). Then, moving the focus down 120 μm, and perform two-dimensional raster scanning again to obtain source PA image 2.

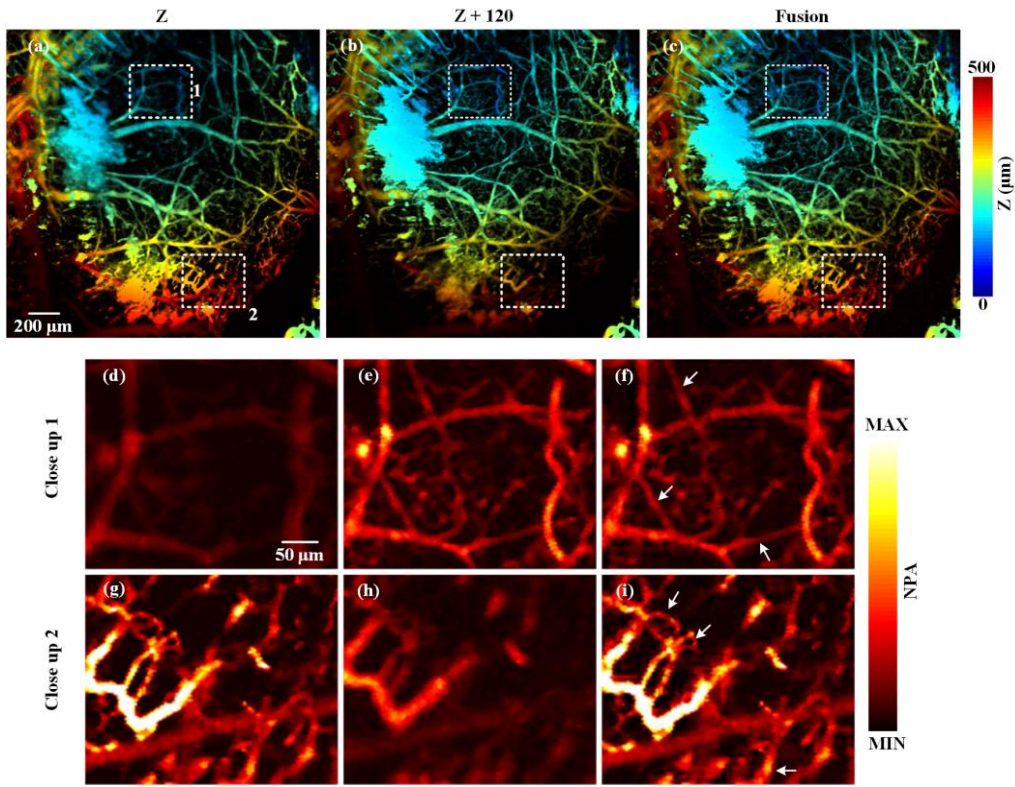

Figure 3. *In vivo* imaging of the mouse cerebral vasculature with intact skull. (a) and (b) are depth-coding max projection (MAP) images obtained when the focus located at z and z +120, respectively. (c) the fused image. (d) - (f) are close-up images of the areas indicated by the rectangles 1 in (a) - (c), respectively; (g) - (i) are close-up images of the areas indicated by the white dashed rectangles in (a) - (c), respectively. The white arrows in (f) and (g) denote the vessels which cannot be resolved in (a) or (b). NPA, normalized amplitude distribution.

Table 2. Performance measure of image fusion of mouse cerebral vasculature.

|  | Entropy | Average gradient | Standard deviation | Edge strength |
|---|---|---|---|---|
| **Image 1 (@ z)** | 5.2083 | 0.7092 | 97.7927 | 7.7993 |
| **Image 2 (@ z +120)** | 5.1791 | 0.5099 | 97.8625 | 5.5383 |
| **Fused image** | 5.9142 | 0.9135 | 136.9177 | 10.0721 |

As shown in Fig. 3, Fig. 3(a) and Fig. 3(b) are depth-coding max projection (MAP) images obtained when the focus located at z and z +120, respectively. Fig. 3(c) is the fused image of Fig. 3(a) and Fig. 3(b). Figs. 3(d) - 3(f) are close-up MAP images of the areas indicated by the white rectangles 1 in Figs. 3(a) - 3(c), respectively; Figs. 3(g) - 3(i) are close-up MAP images of the areas indicated by the white rectangles 2 in Figs. 3(a) - 3(c), respectively. Since the narrow DoF of the OR-PAM, little details can be resolved, only part of the vessels are in focus in Fig. 3(a) and Fig. 3(b). While, the fusion rules enable the system to have a larger depth of field, more microvessels from different depth can be distinguished.

The vessels that are indicated by white arrows can be visualized in fused image, but fuzzy or missing in Fig. 3(a) and Fig. 3(b). Entropy, Average gradient, Mean Square Error (MSE) and Edge strength were calculated to demonstrate the effectiveness of our method quantitatively, as shown in Table 2.

## 4 Conclusion

In summary, by using fusion rules into the PAM system, we developed a computed extended depth of field method for photoacoustic microscope. The wavelet transform coefficients from several source images are fused by electing average of the approximation coefficients and maximum of the detailed coefficients. Images with extended depth of field are obtained by taking inverse wavelet transform. Simulation was performed to demonstrate that this method can extend the depth of field of PAM two times without the sacrifice of lateral resolution. And the in vivo imaging of the mouse cerebral vasculature with intact skull further demonstrates the feasibility of our method. Equipped with a much larger DoF, OR-PAM can be able to further improvement of volumetric imaging speed, which could expand the application of OR- PAM in biomedical researches.

**References**

[1] Hajireza, P., Shi, W., Zemp, R. J., "Label-free in vivo fiber-based optical-resolution photoacoustic microscopy," Opt. Lett. 36(20), 4107–4109 (2011).

[2] Zhang, H. F., Maslov, K., Stoica, G., Wang, L. V., "Functional photoacoustic microscopy for high-resolution and noninvasive in vivo imaging," Nat. Biotechnol. 24(7), 848–851 (2006).

[3] Liu, Y., Zhang, C., Wang, L. V., "Effects of light scattering on optical-resolution photoacoustic microscopy," J. Biomed. Opt. 17(12), 126014 (2012).

[4] Maslov, K., Zhang, H. F., Hu, S., Wang, L. V., "Optical-resolution photoacoustic microscopy for in vivo imaging of single capillaries," Opt. Lett. 33(9), 929–931 (2008).

[5] Chen, J., Lin, R., Wang, H., Meng, J., Zheng, H., Song, L., "Blind-deconvolution optical-resolution photoacoustic microscopy in vivo," Opt. Express 21, 7316–7327 (2013).

[6] Wang, X., Pang, Y., Ku, G., Xie, X., Stoica, G., Wang, L. V., "Noninvasive laser-induced photoacoustic tomography for structural and functional in vivo imaging of the brain," Nat. Biotechnol. 21(7), 803–806 (2003).

[7] Wang, L. V., "Multiscale photoacoustic microscopy and computed tomography," Nat. Photonics 3(9), 503–509 (2009).

[8] Hu, S., Wang, L. V., "Photoacoustic imaging and characterization of the microvasculature," J. Biomed. Opt.15(1), 011101 (2010).

[9] Chen, Z. J., Yang, S. H., Xing, D., "In vivo detection of hemoglobin oxygen saturation and carboxyhemoglobin saturation with multiwavelength photoacoustic microscopy," Opt. Lett. 37(16), 3414–3416 (2012).

[10] Nasiriavanaki, M., Xia, J., Wan, H. L., Bauer, A. Q., Culver, J. P., Wang, L. V., "High-resolution photoacoustic tomography of resting-state functional connectivity in the mouse brain," Proc. Natl. Acad. Sci. U.S.A. 111(1),21–26 (2014).

[11] Li, B., Qin, H., Yang, S., Xing,. D., "In vivo fast variable focus photoacoustic microscopy using an electrically tunable lens," Opt. Express 22(17), 20130 (2014).

[12] Yang, X., Jiang, B., Song, X., Wei, J., Luo, Q., "Fast axial-scanning photoacoustic microscopy using tunable acoustic gradient lens," Opt. Express 25(7), 7349-7357 (2017).

[13] Yang, X., Song, X., Jiang, B., Luo, Q., "Multifocus optical-resolution photoacoustic microscope using ultrafast axial scanning of single laser pulse," Opt. Express 25(23), 28192-28200 (2017).

[14] Yao, J., Wang, L., Yang, J. M., Maslov, K. I., Wong, T. T. W., Li, L., Huang, C. H., Zou, J., Wang, L. V., "High-speed label-free functional photoacoustic microscopy of mouse brain in action," Nat. Methods 12(5), 407–410 (2015).

[15] Yeh, C., Soetikno, B., Hu, S., Maslov, K. I., Wang, L. V., "Microvascular quantification based on contour-scanning photoacoustic microscopy," J. Biomed. Opt. 19(9), 96011 (2014).